# Critical Magnetic Behavior of the Half Heusler Alloy RhCrSi: Monte Carlo Study

**S. IDRISSI[1,*], S. ZITI[2], H. LABRIM[3] and L. BAHMAD[1,**]**

[1] Laboratoire de la Matière Condensée et des Sciences Interdisciplinaires (LaMCScI), Mohammed V University of Rabat, Faculty of Sciences, B.P. 1014 Rabat, Morocco.

[2] Intelligence Artificial and Security of Systems, Mohammed V University of Rabat, Faculty of Sciences, B.P. 1014 Rabat, Morocco.

[3] USM/DERS/Centre National de l'Energie, des Sciences et des Techniques Nucléaires (CNESTEN), Rabat, Morocco.

**Abstract**:

In this paper, we study the critical magnetic properties of the Half Heusler alloy RhCrSi, using Monte Carlo simulations (MCS) under the Metropolis algorithm. In fact, to study this alloy, we apply an Ising model using the MCS simulations, we concentrate only on the magnetic atoms: Rh and Cr. For this purpose, these magnetic atoms are modeled by the spin moments $S=5/2$ for Rh atoms and $\sigma=2$ for Cr atoms, respectively. In addition, we discuss the ground state phase diagrams in different planes corresponding to different physical parameters. On the other hand, for non-null temperature values, we perform the Monte Carlo simulations (MCS) to study the critical behavior of the compound RhCrSi, in the Ising approximation.

Indeed, we present a detailed discussion of the obtained results for the magnetizations as a function of the temperature, the crystal field and the exchange coupling interactions. Additionally, we give the reliance of the basic temperature as an element of precious crystal field when fixing the exchange coupling interactions. To finish this work, we built up and examined the magnetic hysteresis cycles and the relating coercive fields as a part of the external magnetic field.

corresponding authors: *) samiraidrissi2013@gmail.com (S.I.); **) bahmad@fsr.ac.ma (L.B.)

## I. Introduction

In view of their numerous applications, the spintronic has gotten matter of numerous ongoing investigations, in spin-injections, single spin-electron sources. And the latest fields of spintronics applications are the spin-polarized charge carriers (SPCC) [1-5]. This field covers materials, half-Heusler alloys and a gathering of intermetallic materials with the notation 1:1:1. Such alloys have the completely characteristic (SPCC) at the Fermi level. The Primary Half-metallic ferromagnetism has been the compound NiMnSb, which was analyzed by De Groot *et al*. [6].

The Heusler alloys are either half-Heusler alloys XYZ atomic proportions of 1:1:1 or 2:1:1 for X2YZ as full-Heusler compounds [7-13]. The Half-metallic behavior of the Heusler materials is made by the hybridization between the 3d orbitals of the X and Y elements. A novel type of Heusler compounds has attracted in examine intrigue, where X and Y are the fundamental elements of the group [14-26]. On the other hand, Umamaheswari *et al*. [27] found that the ferromagnetic state of the half-Heusler composite LiCaBa is stable than its paramagnetic state. It may be seen that the combination LiXGe (X = CA, Sr or BA) has magnetic moments of one 1.00 $\mu_B$ and fulfills the Salter-Pauling rule $M_{tot} = (8-Z_{tot}) \mu_B$ [28], with $M_{tot}$ and $Z_{tot}$ being the total polarization and the nuclear number, respectively.

In addition, Sun *et al*. [29] have discussed the origin of magnetism in the compound $LiCsN_2$ brought by the '2p' orbital from the N atoms. Generally, in the Heusler alloys, in which X and Y are fundamental elements, are called 'sp' or 'd0' half-metals, and have essentially minimal magnetic moments. Which makes them continuously precious in the spintronic applications. On the other hand, Cherrid *et al*. [30] studied the different physical properties of the Full Heusler $Cs_2CrGe$ compound, the electronic, the mechanical, the magnetic properties and revealed half-metallic behavior and stable mechanical properties by the GGA approximation of this alloy. Other most research has approved the HM materials of the Heusler structures, see Refs. [31-41]. Recently, the half-metallicity was passed in combinations of Heusler or half-Heusler except the transition metals, such as RbSrX2 (X = C, N and O), KCaX2 (X = C, N and O), RbSrZ. (Z = C, Si or Ge) and CsBaX2 (X = C, N or O) [42-45]. More recently, the half-Heusler RhCrSi alloy has been treated theoretically by Ab-initio calculations to study its structural, elastic, mechanical, electronic, magnetic and optical characteristics [46].

In this work, we study the magnetic properties and critical behavior of the half-Heusler RhCrSi compound. In order to investigate its total magnetizations and susceptibilities as a subroutine of different physical parameters. We have applied Hamiltonian using Ising model under Metropolis algorithm [47-48]. In fact, we illustrate such behavior as a function of temperature, exchange coupling interactions, crystal field and external magnetic field. The second part is dedicated to a theoretical study of the Half Heusler alloy. In Part III, we illustrate the results obtained by Monte Carlo Simulation (MCS). We conclude in Section IV by describing the overall results obtained for the half-Heusler RhCrSi alloy.

## II. Theoretical model of the Half Heusler Alloy RhCrSi

The Half Heusler RhCrSi alloy crystallizes in the three-dimensional structure and belongs to the space group F43m (No. 216). The Wyckoff positions are occupied by X atoms in the non-equivalent 4c (0.25, 0.25, 0.25), the Y and Z atoms are attached to the 4d (0.75, 0.75, 0, 75), 4a (0, 0, 0), respectively. We deduce the location of the atoms of the studied alloy are depicted from the Ref. [50].

The crystal structure of the half-Heusler RhCrSi alloy is illustrated in Fig.1 (a). Also the structure of this alloy describing only magnetic atoms and the different exchange coupling interactions is presented in Fig.1 (b), in order to build these geometries, we use the VESTA package [49].

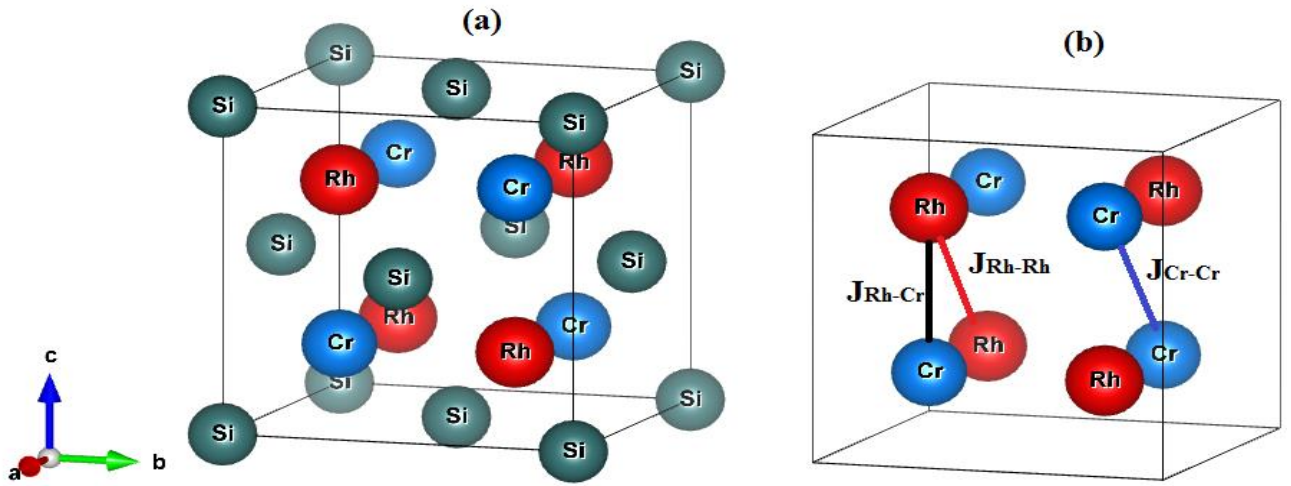

*Fig.1: Crystal structure of the half-Heusler alloy RhCrSi in (a); Structure describing only magnetic atoms and the different exchange coupling interactions in (b).*

The Hamiltonian describing the studied system is modeled by an Ising model expressed as follows:

$$\mathcal{H} = -J_{Rh-Rh} \sum_{i,j} S_i S_j - J_{Cr-Cr} \sum_{k,l} \sigma_k \sigma_l - J_{Rh-Cr} \sum_{i,k} S_i \sigma_k - H \sum_i (S_i + \sigma_i) - \Delta_\sigma \sum_i \sigma_i^2 - \Delta_S \sum_i S_i^2$$

(1)

Where: $S_i = \pm 5/2, \pm 3/2, \pm 1/2$: corresponding to the Rh atoms. The Cr atoms are modeled by $\sigma_i = \pm 2, \pm 1$ and 0.

$J_{Rh-Rh}$, $J_{Cr-Cr}$ and $J_{Rh-Cr}$ are the exchange interactions between Rh-Rh, Cr-Cr and Rh-Cr atoms, respectively. In this work, we are limited only to the first nearest neighbor atoms. $\Delta_S$, $\Delta_\sigma$: are the crystal fields acting on the Rh and Cr ions, respectively. For simplicity, in all this work, we will assume that the internal field is constant: $\Delta = \Delta_S = \Delta_\sigma$.

The total energy per site of the Half Heusler RhCrSi is calculated as:

$$E_T = \frac{1}{N} <\mathcal{H}> \quad (2)$$

Where $N = N_S + N_\sigma$, with $N_S, N_\sigma$ are the number of S and σ atoms that belong to the super cell unit, respectively.

The total magnetization is obtained by:

$$m = \frac{N_\sigma m_\sigma + N_S m_S}{N} \quad (3)$$

With $m_S = \frac{1}{N_S} \sum_i S_i$ and $m_\sigma = \frac{1}{N_\sigma} \sum_i \sigma_i$

The total susceptibility is given by:

$$\chi = \frac{<M_T^2> - <M_T>^2}{k_B T} \quad (4)$$

The partial susceptibilities are: $\chi_S = \frac{<M_S^2> - <M_S>^2}{K_B T}$ and $\chi_\sigma = \frac{<M_\sigma^2> - <M_\sigma>^2}{K_B T}$, respectively.

The specific heat are calculated by the following expressions:

$$C_v = \beta^2 (<E_T^2> - <E_T>^2) \quad (5)$$

Where $\beta = \frac{1}{k_B T}$: T is the absolute temperature and $k_B$ represents the Boltzmann constant. It is fixed to its unit value, in all the following.

## III. Monte Carlo Simulation (MCS) of the Half Heusler RhCrSi alloy

Magnetic properties of the Monte Carlo simulation (MCS) based on the Metropolis algorithm of the Half Heusler RhCrSi alloy. Our calculations are based on the equation Hamiltonian. (1), where free boundary conditions are used along the super-cell unit of size N = 5x5x5. We perform $10^5$ Monte Carlo steps for all spin configurations. All sites in the system are scanned and an attempt to flip to a rotation is created. We accept or deny the flips of each turn, according to Boltzmann's statistics. Total partial magnetizations, total partial magnetic susceptibility, system energy and specific heat are calculated at the counterweight of the regime. Via Equation (Eq.2), for each iteration, we estimate the internal energy per site. The error bars are counted with a long-haired knife method [51]. The error bars are smaller than the symbol sizes of different shapes of magnetic properties. More technical details of the used Monte Carlo method have been outlined in some our works [52-58]. While other materials have been investigated by using DFT and TDDFT methods [59-65].

**III.1. Discussion of the ground state phases**

In this part, we analyze and discuss the ground states, phases of the Heusler alloy RhCrSi at temperature (T=0 K). We use Eq. (1) to assume the energy of each conformation of this mixed system (S, σ). The more stable configuration corresponds to the lower limit of the energy, estimated by the Eq. (1), for the all possible configurations: $(2S+1)*(2σ+1) = 30$, with S=5/2 and σ=2.

The stable phases in the ground state phase diagrams are presented in Figs. 2 (a), 2 (b) and 2 (c) for $J_{Rh-Rh}=J_{Cr-Cr}=J_{Rh-Cr}=1$. In fact, Fig.2 (a) shows the stable configurations in the plane (D, H) for $J_{Cr-Cr}=1$. From this figure it is found that only 8 stable ones are present in this figure, namely: (2.5, 2.0), (0.5, 0.0), (1.5, 1.0), (0.5, 1.0) and (0.5, 0.0) and their opposites symmetrical regarding the axis H=0. In fact a perfect symmetry is found in this figure concerning the stable phases. In the plane ($J_{Rh-Cr}$, H), the stable phases are reported in Fig.2 (b) for Δ=0 and $J_{Rh-Rh}=J_{Cr-Cr}=1$. As in Fig.2 (a), this figure shows a perfect symmetry regarding the axis H=0. The stable phases in the plane ($J_{Rh-Cr}$, Δ) are presented in Fig.2 (c) for H=0 and $J_{Rh-Rh}=J_{Cr-Cr}=1$. In the region with Δ<-5, the all stable phases are present. While for Δ>-5, the only stable phases are: (-2.5, -2.0), (-1.5, -1.0) and (-0.5, 0.0), see Fig.2.c.

The ground state phase diagrams of the studied compound are presented in Figs.3 (a), 3 (b) and 3 (c) for Δ=1 and H=1. In fact, Fig.3 (a) represents the stable configurations in the plane ($J_{Rh-Cr}$, $J_{Rh-Rh}$) for $J_{Cr-Cr}= 1$. From this figure, it is found that from the 6x5=30 possible configurations, only 10 stable ones are present in this figure, namely: (-1.5, -2.0), (-0.5, -2.0), (0.5, -2.0), (-2.5, -2.0), (-1.5, -2.0), (-2.5, -2.0), (-2.5, -2.0), (-2.5, -2.0); (1.5, -2.0) and (-1.5, -2.0).

The phases with maximum spin moment are to be stable for $J_{Rh-Rh}$ with positive values. In the plane ($J_{Cr-Cr}$, $J_{Rh-Rh}$) we plot Fig.3 (b) showing the stable phases for $J_{Rh-Cr}= 1$. From this figure, it is seen that only four phases are stable, namely: (-2.5, -2.0), (-0.5, -2.0), (-0.5, 0.0) and (-2.5, 0.0). Some of these phases are reappearing in different areas, for example the phase (-0.5, 0.0) is stable for whatever values of $J_{Rh-Rh}$ and any negative values of $J_{Cr-Cr}$. In the plane ($J_{Cr-Cr}$, $J_{Rh-Cr}$), we describe the stable phases in Fig.3 (c) for $J_{Rh-Rh}=1$. It is found that six phases are stable in this figure, namely: (-2.5, -2.0), (-2.5,

2.0), (-2.5, 0.0), (-2.5, 1.0), (-2.5, -1.0) and (-2.5, 1.0). The phases with maximum spin moment are to be stable for positive values of the exchange coupling interaction $J_{Cr-Cr}$.

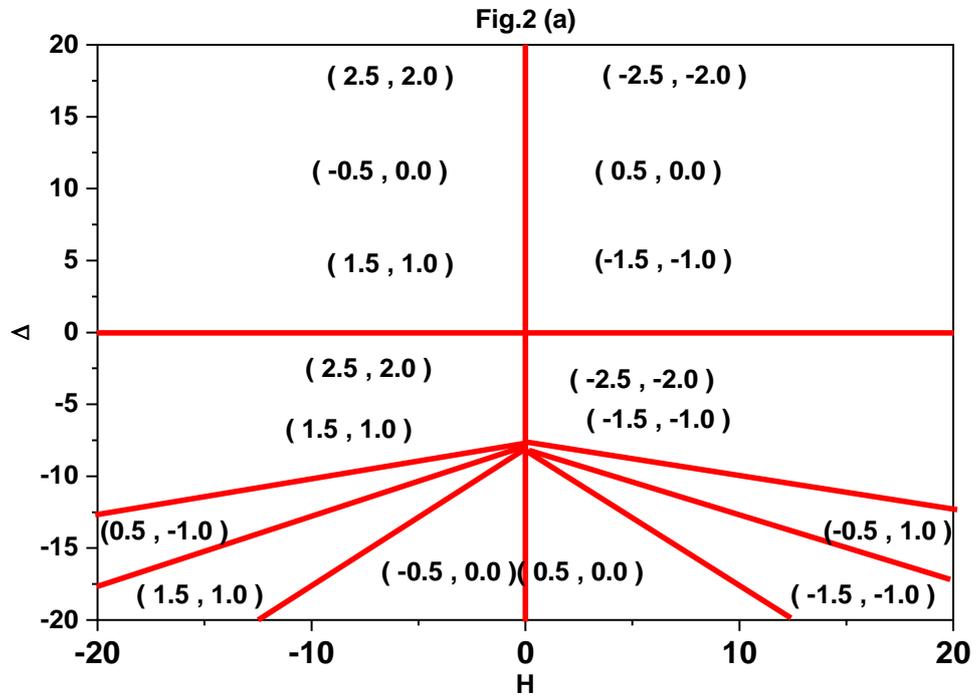

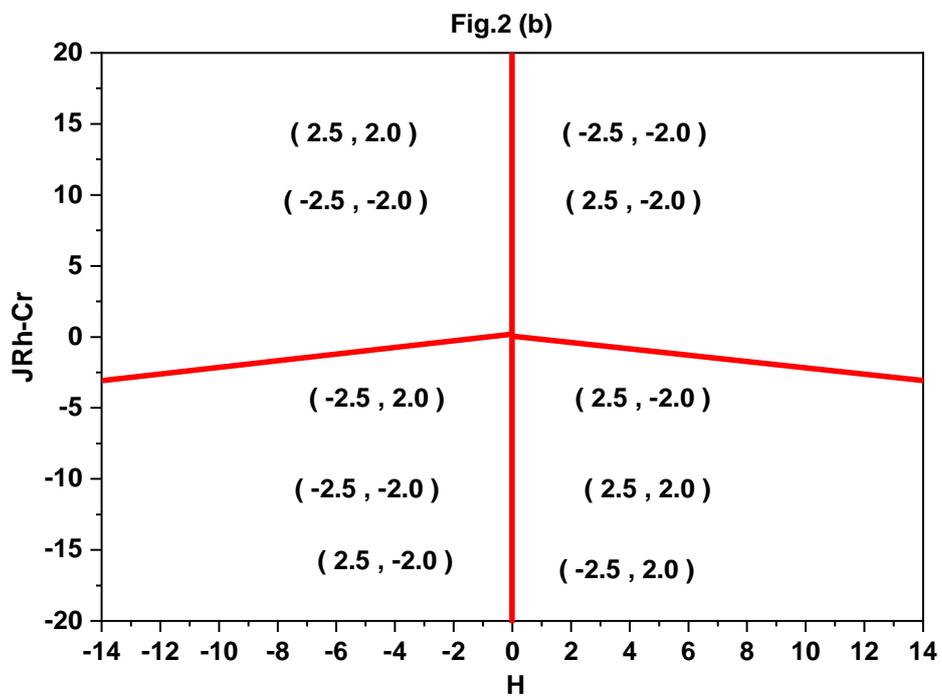

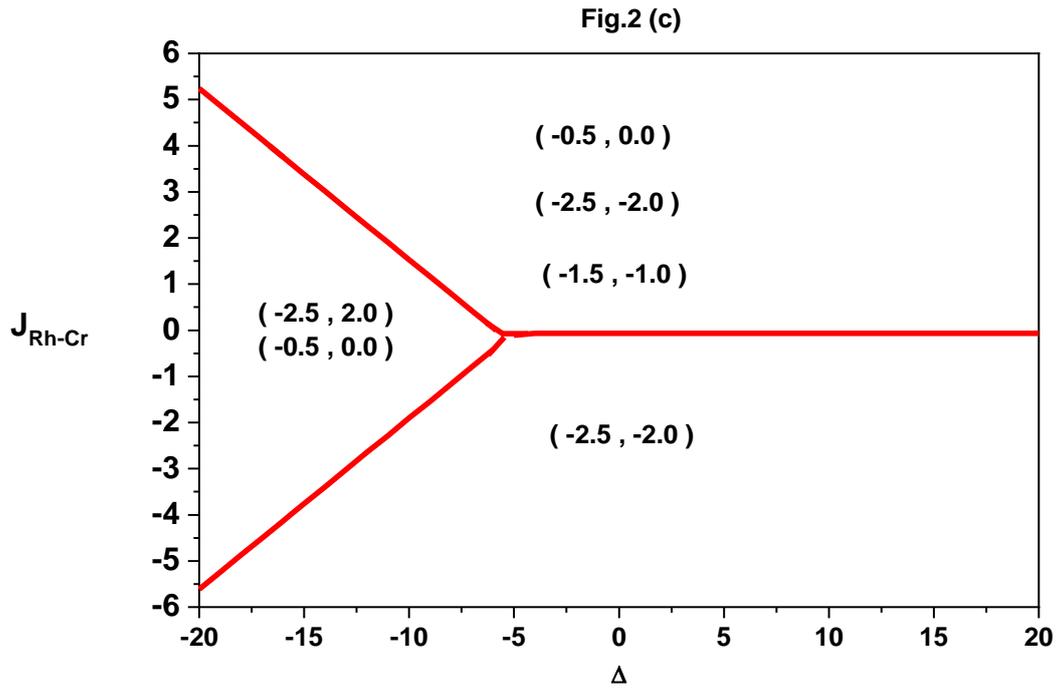

*Fig.2: Ground state phase diagrams: in (a) the plane ($\Delta$, H) for $J_{Rh-Rh}=J_{Cr-Cr}=J_{Rh-Cr}=1$, in (b) the plane ($J_{Rh-Cr}$, H) for $\Delta=0$ and $J_{Rh-Rh}=J_{Cr-Cr}=1$ and in (c) the plane ($J_{Rh-Cr}$, $\Delta$) for H=0 and $J_{Rh-Rh}=J_{Cr-Cr}=1$.*

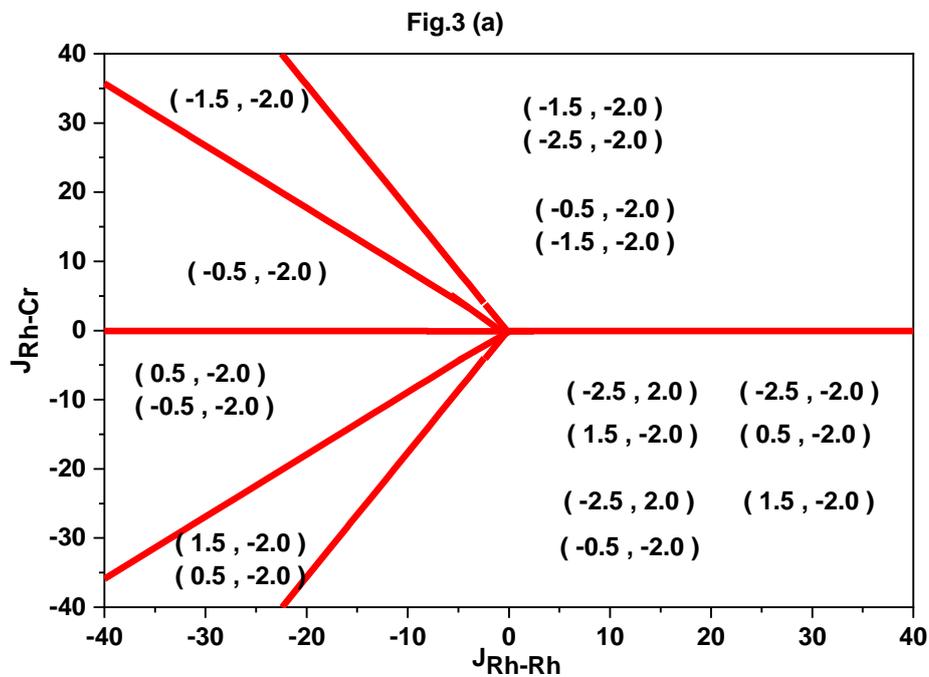

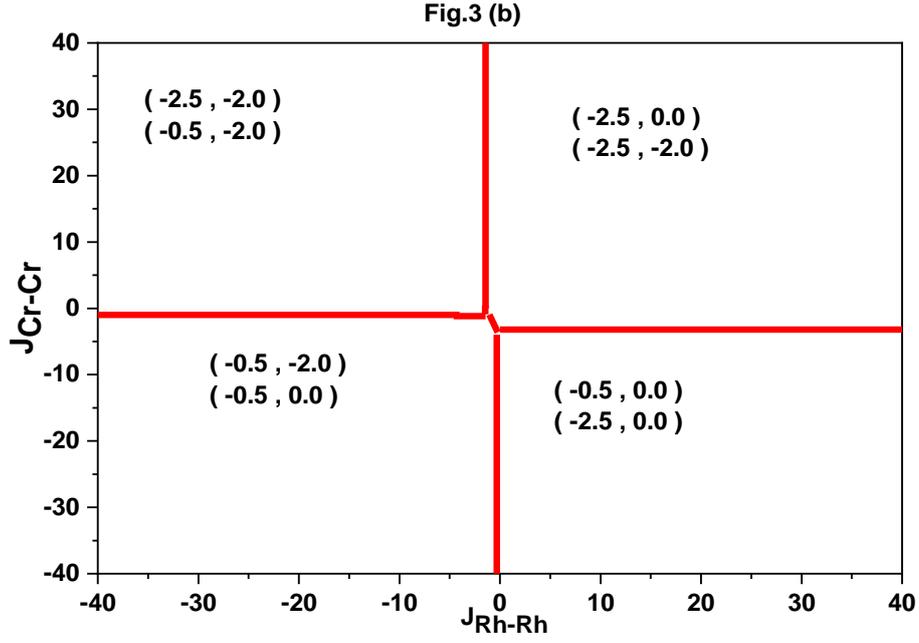

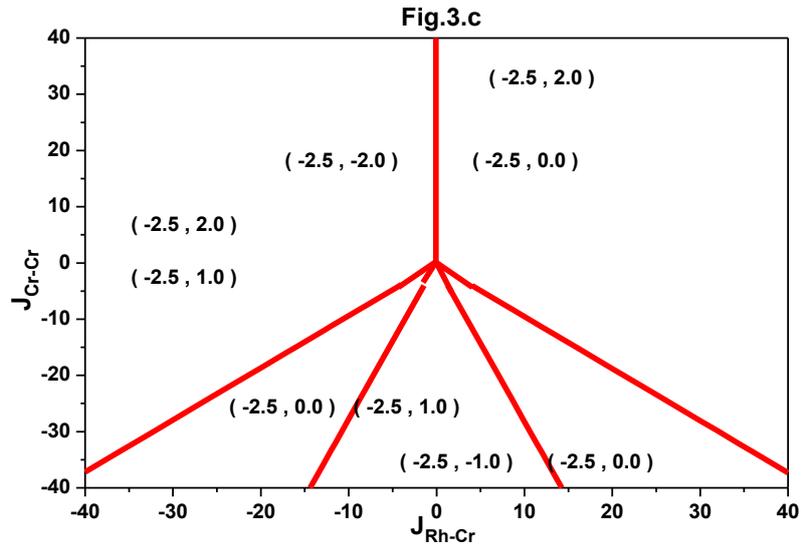

*Fig.3: Ground state phase diagrams for Δ=1 and H=1: in (a) the plane ($J_{Rh-Cr}$, $J_{Rh-Rh}$) for $J_{Cr-Cr}= 1$, in (b) the plane ($J_{Cr-Cr}$, $J_{Rh-Rh}$) for $J_{Rh-Cr}= 1$ and in (c) the plane ($J_{Cr-Cr}$, $J_{Rh-Cr}$) for $J_{Rh-Rh}= 1$.*

### III.2. Results of the Monte Carlo method

In order to study the transition behavior of the Half Heusler alloy RhCrSi, we use the Hamiltonian of the Eq. (1) to simulate the conduct of this compound when varying the temperature, the crystal field, the different exchange coupling interactions and the external magnetic field.

We start by exploring the behavior of the total magnetizations as a function of the temperature, In order to illustrate the thermal behavior of the magnetic proprieties of the studied system, we report in Figs.4 (a), 4 (b) and 4 (c), the obtained results for the specific values: H=0, Δ=0 and $J_{Rh-Rh}=J_{Cr-Cr}=J_{Rh-Cr}$=1. Fig.4 (a) represents the total and partial magnetizations of the Half Heusler RhCrSi compound. At very low temperature values, we obtain the ground state spin moment values. Namely, S=5/2 for the Rh atoms, σ=2 for the Cr atoms and m=(S+σ)/2=2.25. The corresponding total and partial susceptibilities are provided in Fig.4 (b). In connection with Fig.4 (a), the peak of both the susceptibility, plotted in Fig.4 (b) and the specific heat specific heat illustrated in specific heat; corresponds to the value the transition temperature.

The effect of varying different values of the crystal field on the behavior of the transition temperature is reported in Figs.5 (a) and 5 (b) for H=0 and $J_{Rh-Rh}=J_{Cr-Cr}=J_{Rh-Cr}$=1. The peak of each susceptibility curve corresponds to the value of the transition temperature. These figures are plotted for Δ=3, Δ=5 and Δ=10 in Fig.5 (a) and for Δ=15, Δ=20 and Δ=25 in Fig.5 (b). From these figures, the increasing crystal field effect is to increase the transition temperature. Such finding is confirmed in Fig.5 (c). A linear behavior is presented for the transition temperature as a function of the crystal field.

The effect of varying different exchange coupling interactions on the behavior of the transition temperature is reported in Figs.6 (a) and 6 (b) for H=Δ=1. In fact the peak of each susceptibility curve corresponds to the value of the transition temperature. These figures are plotted for J=1, J=5 and J=10 in Fig.6 (a) and J=15, J=20 and J=25 in Fig.6 (b), with $J=J_{Rh-Rh}=J_{Cr-Cr}=J_{Rh-Cr}$. From these figures the increasing exchange coupling interactions effect is to increase the transition temperature. This result is confirmed in Fig.6 (c), where an almost linear behavior is presented for the transition temperature when varying the exchange coupling interactions.

To examine the effect of varying the exchange coupling interactions on the behavior of the total magnetizations of the Half Heusler alloy RhCrSi, we provide in Figs.7 (a), 7 (b) and 7 (c), the corresponding results for the fixed temperature value T=20 K and selected values of the crystal field: Δ=-2, Δ=0 and Δ=+2, in the absence of the external magnetic field.

In Fig.7 (a), we present the variation of the total magnetizations when varying the exchange coupling interaction $J_{Rh-Rh}$ for the specific value of the couplings $J_{Cr-Cr}=J_{Rh-Cr}$=1. From this figure, it is found that the total magnetizations undergo a second order transition when increasing the $J_{Rh-Rh}$ parameter. The magnetization saturation follows the sign of the crystal field. This saturation is positive for Δ=0 and Δ=+2 and negative for Δ=-2.

To examine the effect of varying the parameter $J_{Rh-Cr}$ on the behavior of the total magnetizations, we illustrate in Fig.7 (b), the obtained results for a fixed value of the couplings $J_{Rh-Rh}= J_{Cr-Cr}=1$. From this pattern, it is determined that the total magnetizations undergo a first order transition when increasing the parameter $J_{Rh-Cr}$. For negative values of the exchange coupling interaction $J_{Rh-Cr}$ the variation of the crystal field does not affect total magnetizations, see Fig.7 (b).

In Fig.7 (c), we provide the variation of the total magnetizations as a function of the exchange coupling interaction $J_{Cr-Cr}$ for a fixed value of the couplings $J_{Rh-Rh}= J_{Rh-Cr}=1$. From this figure, it is clear that the total magnetizations undergo a second order transition when increasing the parameter $J_{Cr-Cr}$. The magnetization saturation follows the sign of the crystal field. This saturation is negative for $\Delta=-2$ and positive for $\Delta=0$ and $\Delta=+2$.

The hysteresis loops of the half Heusler RhCrSi alloy, are plotted in Figs.8 (a), 8 (b) and 8 (c) corresponding to the variation of temperature, the different exchange coupling interactions and the crystal field, respectively. In fact, we present in Fig.8 (a) the obtained results for $\Delta=0$, $J_{Rh-Rh}=J_{Cr-Cr}=J_{Rh-Cr}=1$ and selected values of temperature: T=5 K, 15 K, 22 K and 100 K. When exploring this figure, we found that the increasing temperature effect is not just to decrease the surface of the hysteresis loops, but also to decrease the saturation value of the total magnetizations.

The effect of varying the exchange coupling interactions on the conduct of the hysteresis cycles, for fixed values of the crystal field ($\Delta=0$) and temperature (T=20 K), is given in Fig.8.b. This is outlined for the specific values of the exchange coupling interactions for the following cases:

(J1) corresponding to: $J_{Rh-Rh}= J_{Rh-Cr}= J_{Cr-Cr}=2$,

(J2) with the values: $J_{Rh-Rh}= ,J_{Rh-Cr}= J_{Cr-Cr}=1$,

(J3) for the case: $J_{Rh-Rh}= ,J_{Rh-Cr}=-1$, $J_{Cr-Cr}=1$,

(J4) for the values: $J_{Rh-Rh}=-1 ,J_{Rh-Cr}=1, J_{Cr-Cr}=-1$.

From the Fig. 8 (a), it is found that the surface of the hysteresis loops is proportional to the amplitude of the exchange coupling interactions when these parameters are taking positive values, see Fig. 8 (b) for $J_{Rh-Rh}= J_{Rh-Cr}= J_{Cr-Cr}=2$ and $J_{Rh-Rh}= J_{Rh-Cr}= J_{Cr-Cr}=1$. On the other hand the appearance of the steps in the hysteresis cycles is due to the antiferromagnetic behavior of the Half Heusler RhCrSi alloy for the specific values of the exchange coupling interactions: $J_{Rh-Rh}= J_{Rh-Cr}=-1, J_{Cr-Cr}=1$ and $J_{Rh-Rh}=-1, J_{Rh-Cr}=1, J_{Cr-Cr}=-1$. In order to inspect the effect of varying the crystal field on the hysteresis cycles, we provide in Fig.8 (c), the obtained results for the temperature T=25 K and fixed values of the couplings $J_{Rh-}$

$_{Rh}$=J$_{Cr-Cr}$=J$_{Rh-Cr}$=1. This pattern is plotted for the selected values of crystal field Δ= -2, Δ= 0, Δ=2 and Δ=4. From this figure, it is clear that the increasing crystal field effect is to increase the surface of the hysteresis loops.

To complete this survey, we provide in Fig. 9 the variation of the coercive field as a part of the crystal field for T=25 K and J$_{Rh-Rh}$=J$_{Cr-Cr}$=J$_{Rh-Cr}$=1. From this frame, it is set up that the coercive field undergoes two behaviors: for Δ <4, this parameter presents an almost linear variation. While for Δ >4, the coercive field is not affected by the increasing the values of the crystal field.

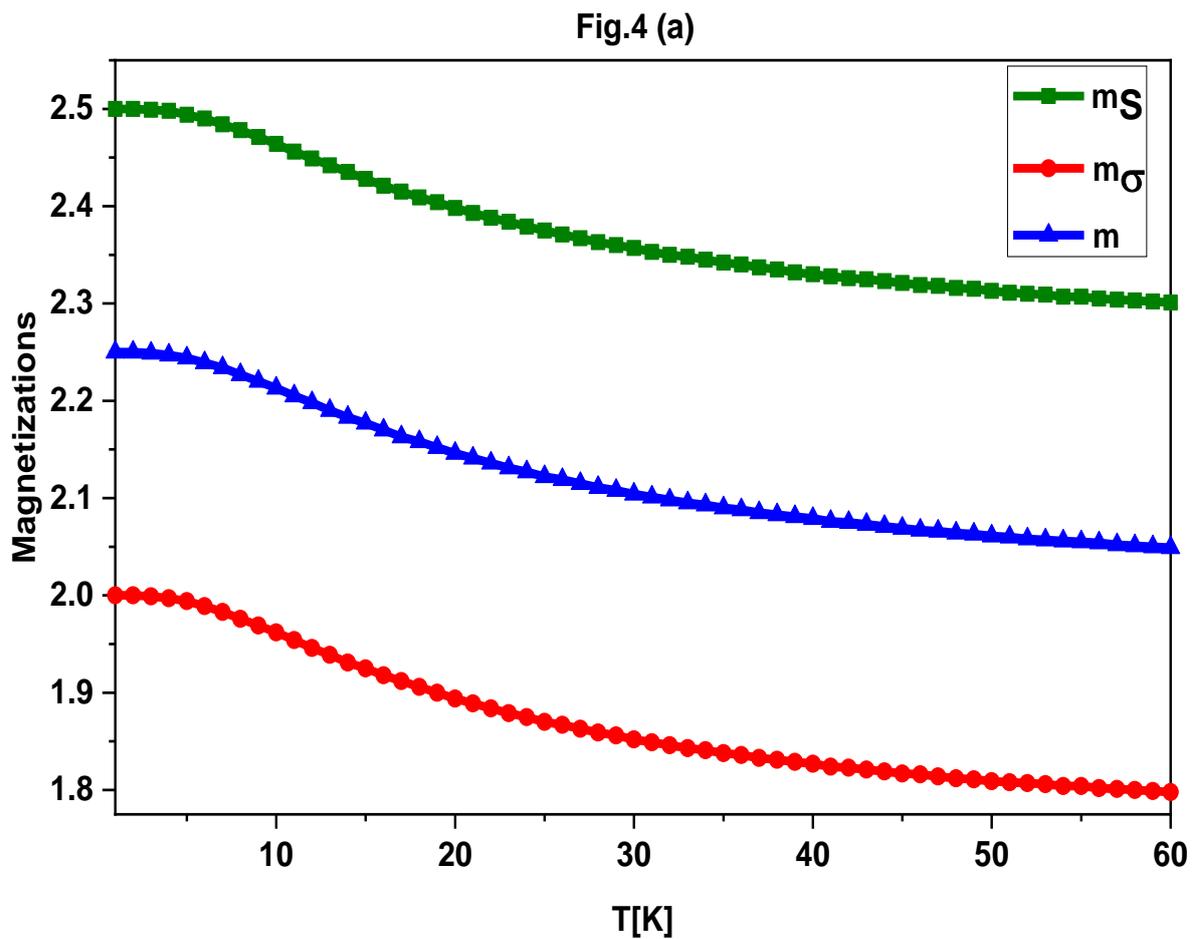

Fig.4 (a)

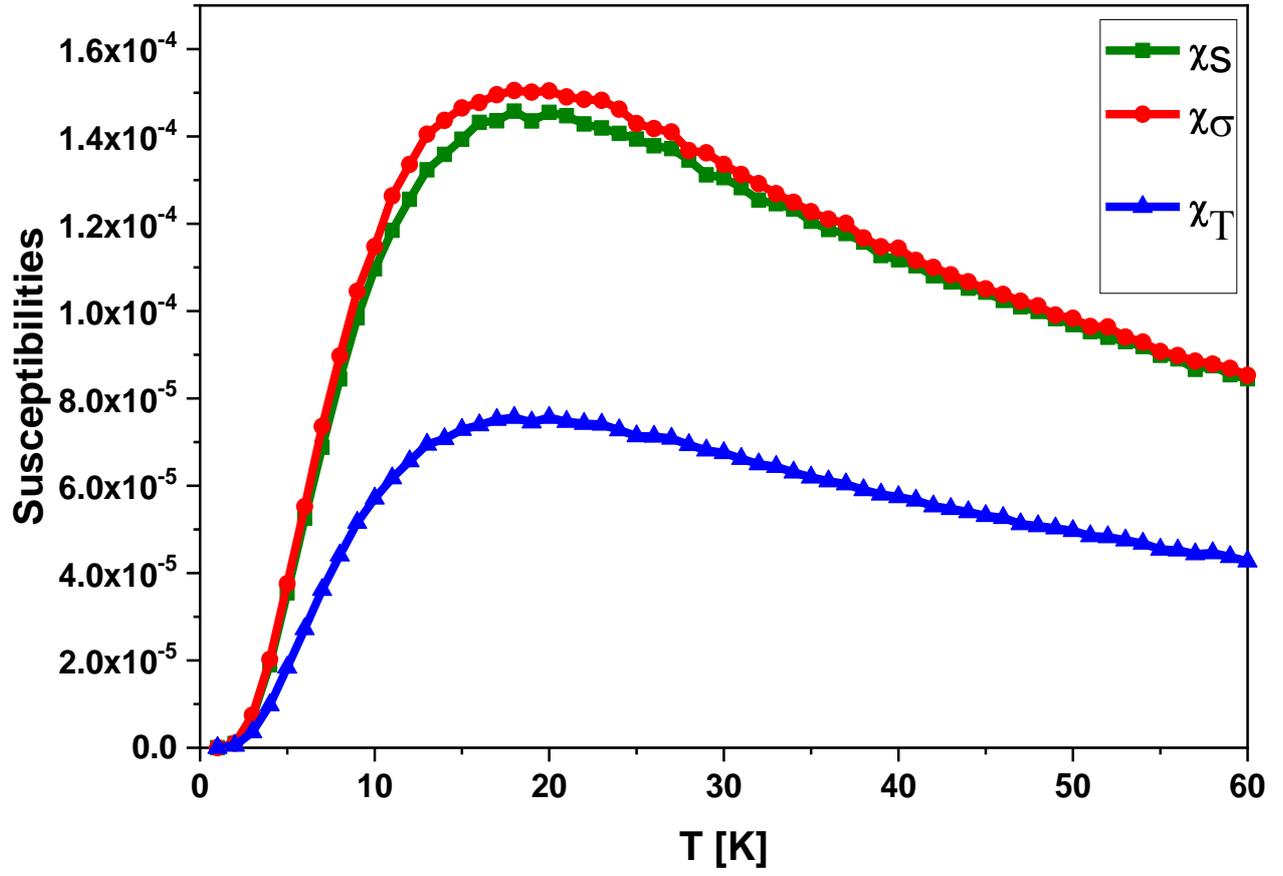

Fig.4 (b)

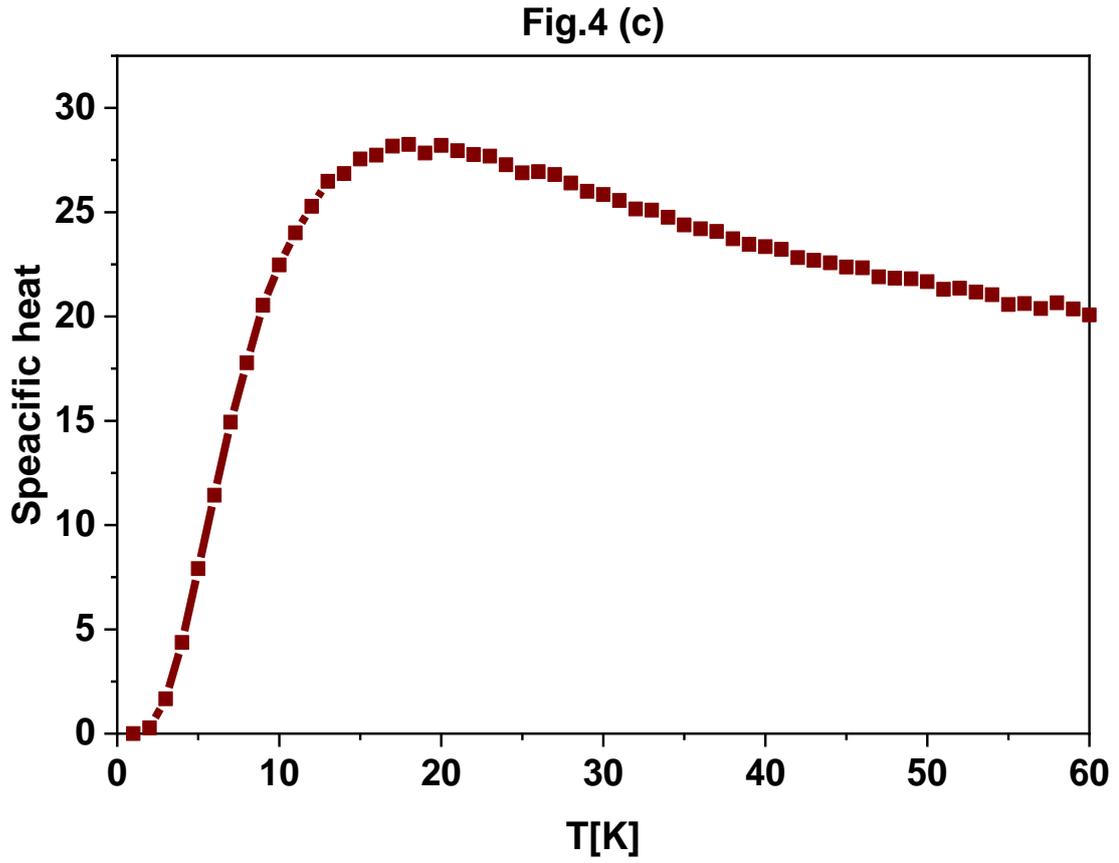

*Fig.4: Thermal behavior of magnetic proprieties of the studied system for H=0, Δ=0 and J$_{Rh-Rh}$=J$_{Cr-Cr}$=J$_{Rh-Cr}$=1. In (a) the total and partial magnetizations. In (b) the total and partial susceptibilities. In (c) the specific heat.*

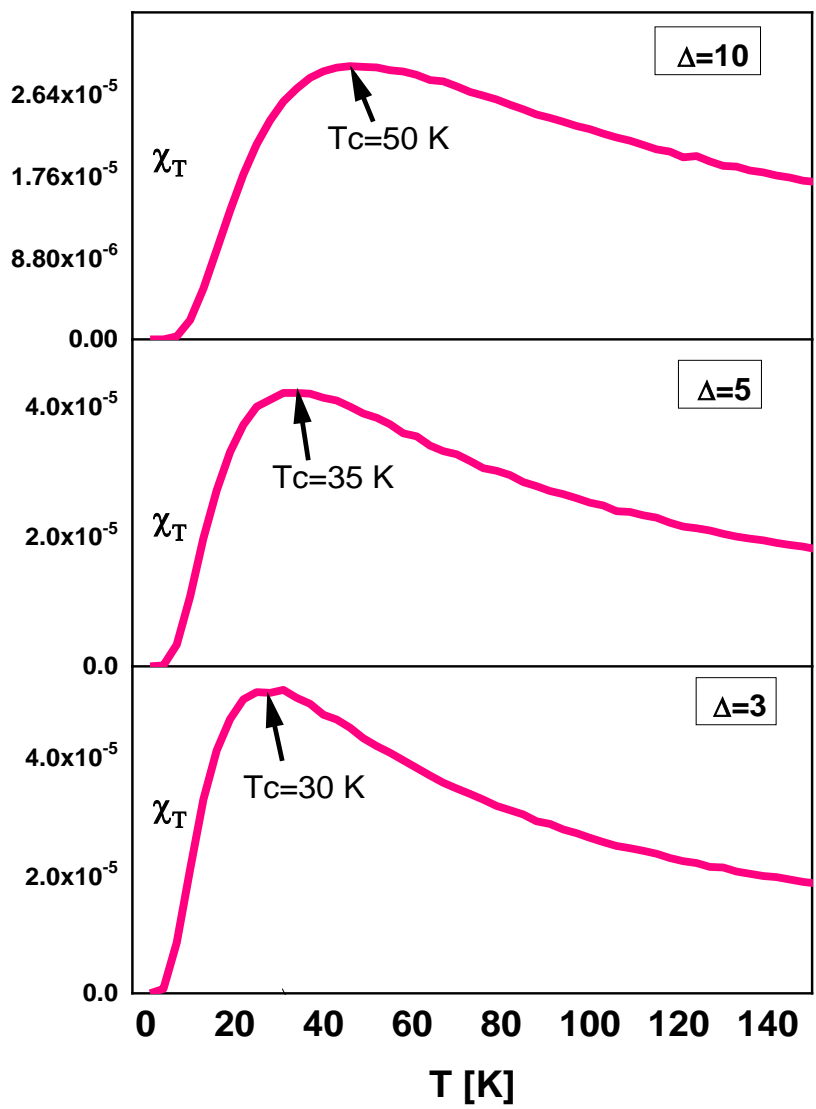

**Fig.5 (a)**

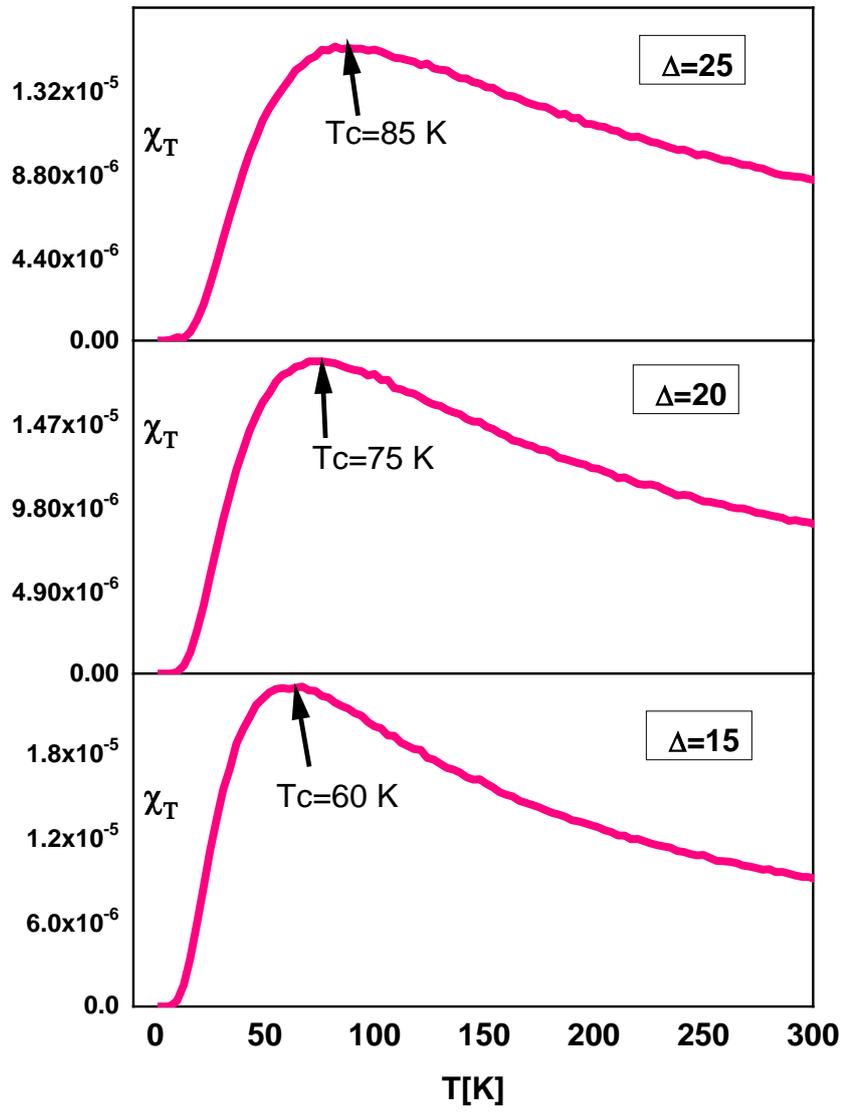

Fig.5 (b)

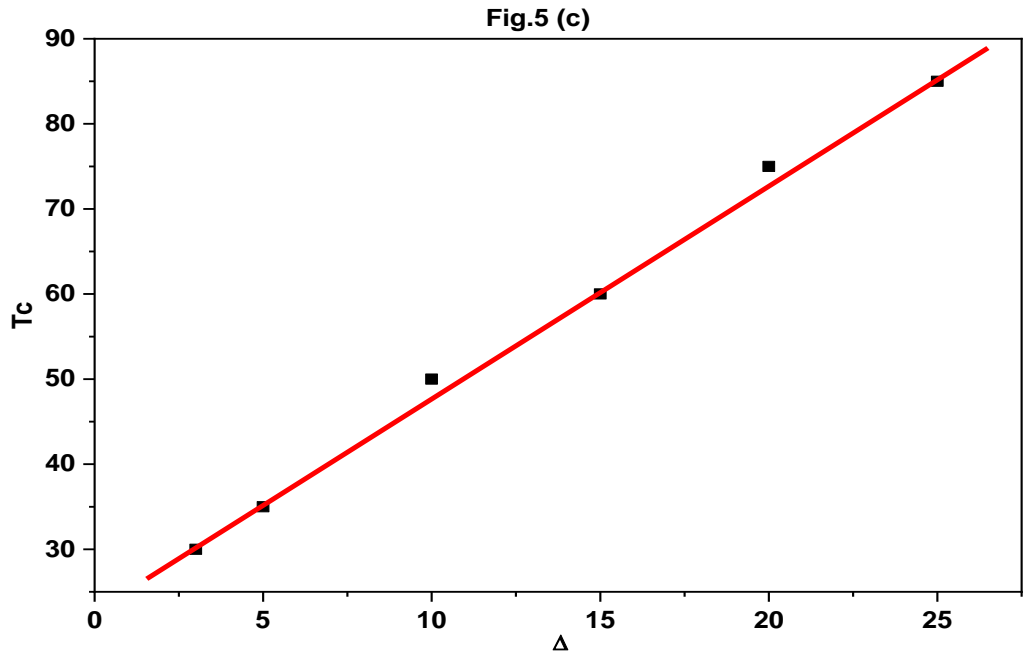

*Fig.5: Thermal behavior of the studied system for H=0 and $J_{Rh-Rh}=J_{Cr-Cr}=J_{Rh-Cr}=1$. In (a) for $\Delta=3$, $\Delta=5$ and $\Delta=10$. In (b) for $\Delta=15$, $\Delta=20$ and $\Delta=25$. In (c) the transition temperature as a function of the crystal field.*

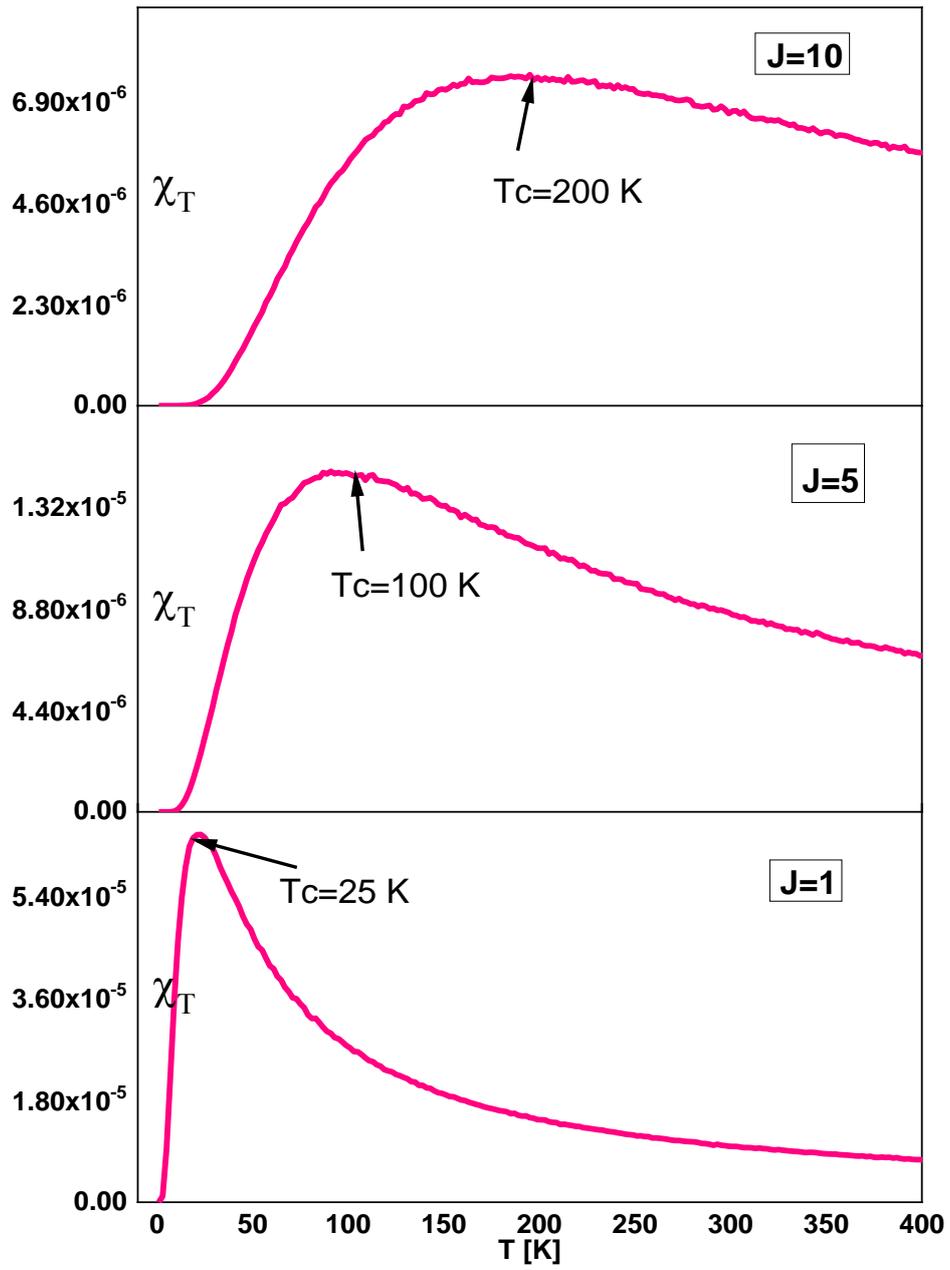

**Fig.6 (a)**

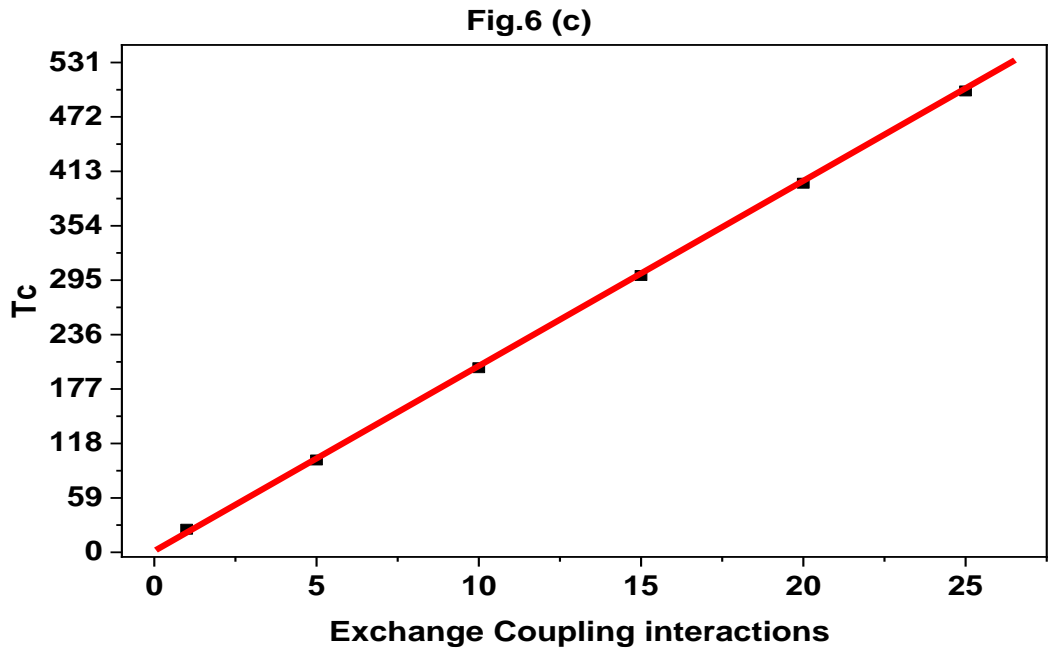

*Fig.6: Thermal behavior of susceptibility of the studied system for H=Δ=1 and J=J$_{Rh-Rh}$=J$_{Cr-Cr}$=J$_{Rh-Cr}$=1. In (a) for J=1, J =5 and J= 10. In (b) for J=15, J =20 and J= 25. In (c) the transition temperature as a function of the exchange coupling interactions.*

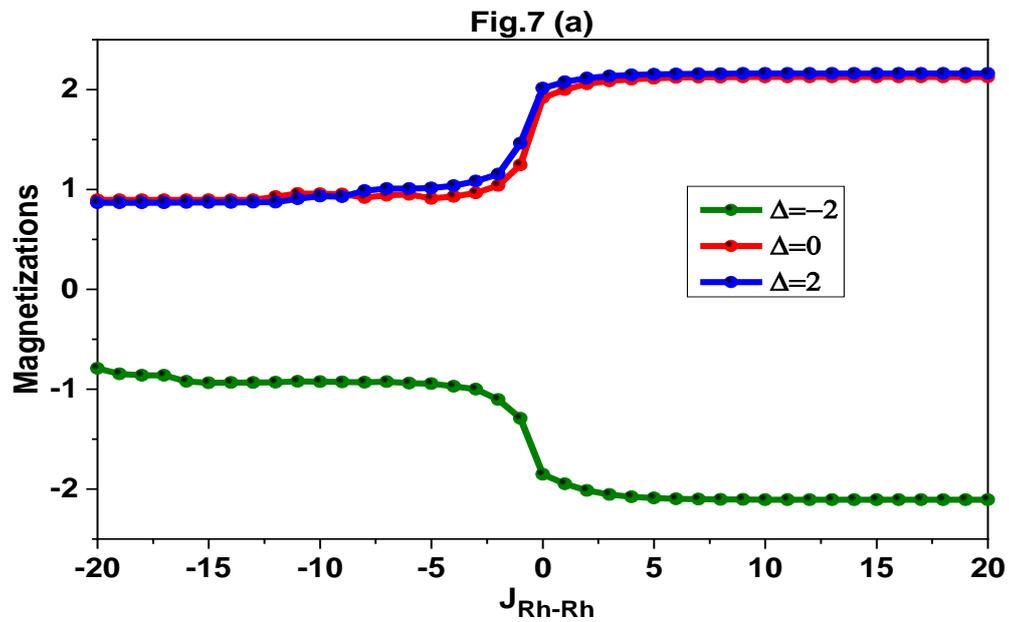

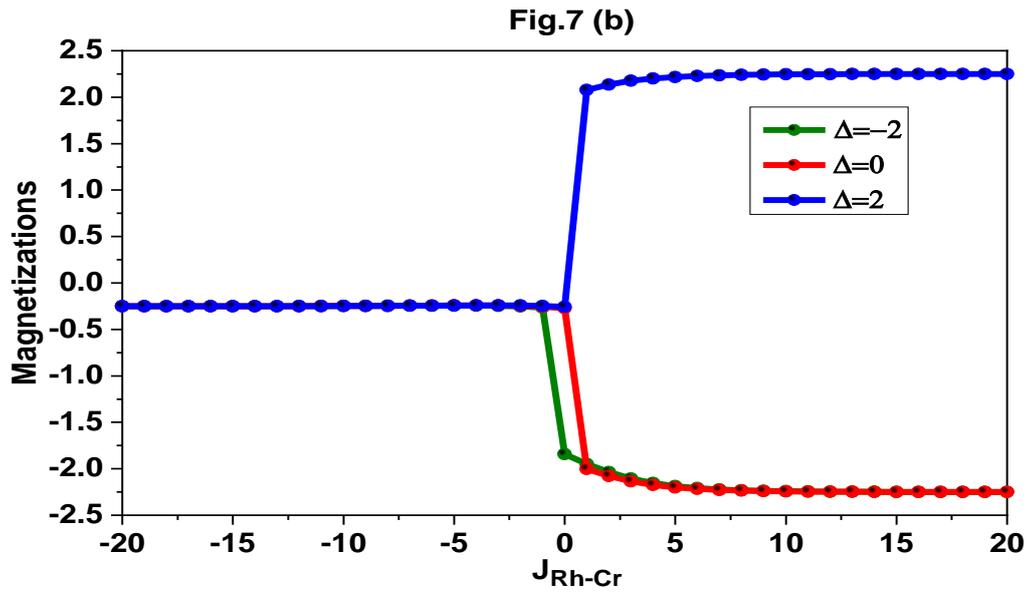

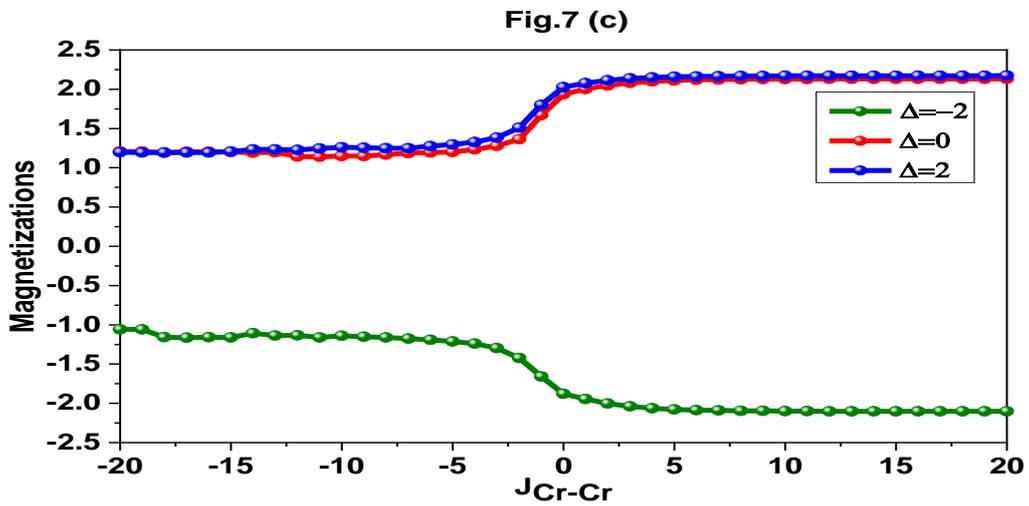

*Fig.7: Magnetization behavior as a function of exchange coupling interactions for T=20 K, H=0 and selected values of crystal field Δ=-2, Δ=0 and Δ=+2: In (a) $J_{Rh-Rh}$ for $J_{Cr-Cr}=J_{Rh-Cr}=1$. In (b) $J_{Rh-Cr}$ for $J_{Rh-Rh}= J_{Cr-Cr}=1$. In (c) $J_{Cr-Cr}$ for $J_{Rh-Rh}= J_{Rh-Cr}=1$.*

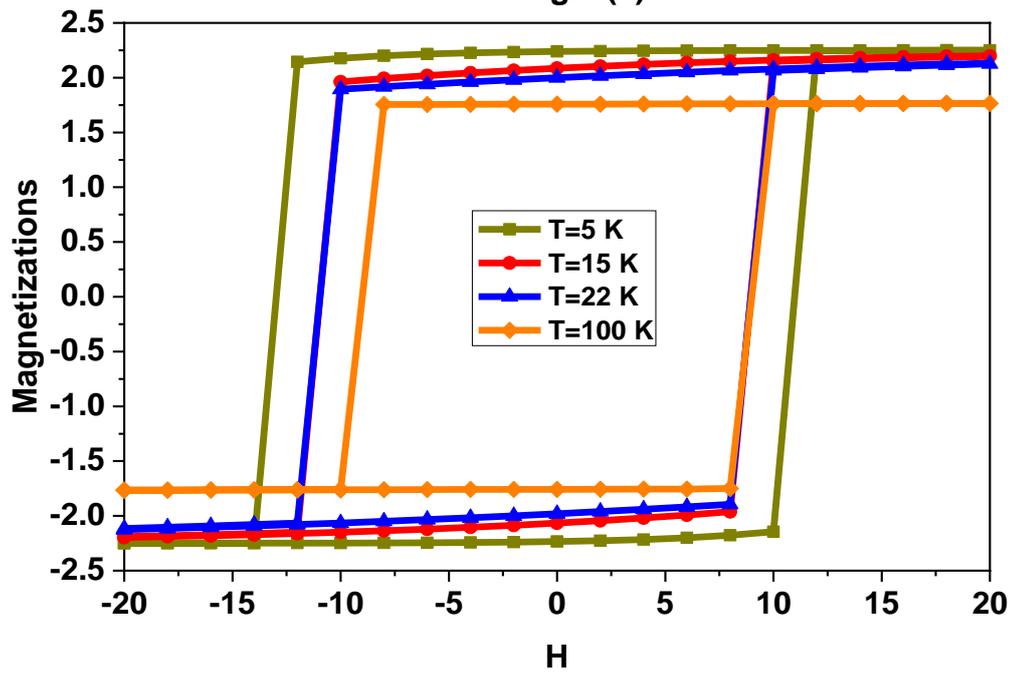

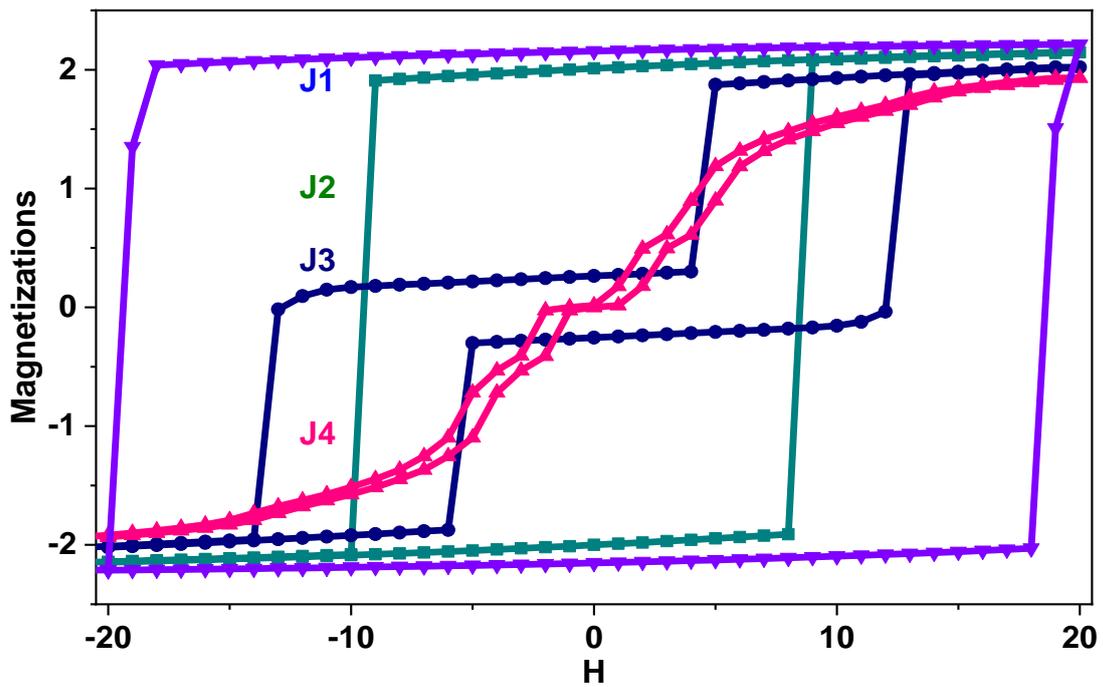

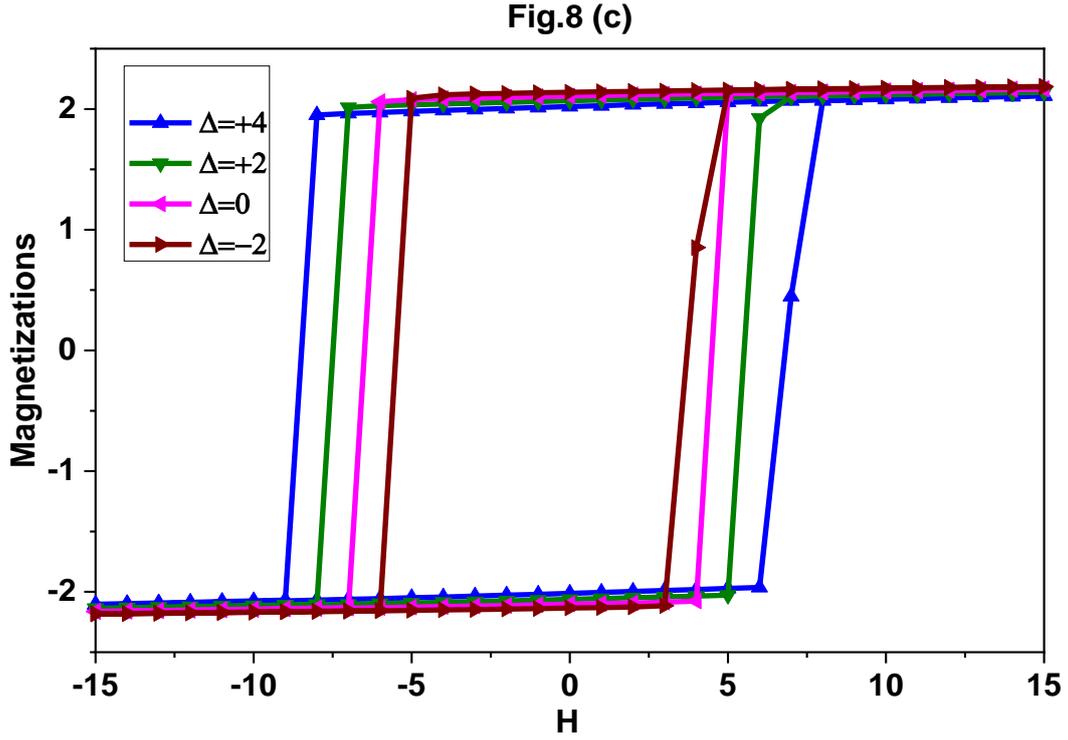

*Fig.8: The hysteresis loops of the half Heusler alloy RhCrSi. In (a) for $\Delta=0$, $J_{Rh-Rh}=J_{Cr-Cr}=J_{Rh-Cr}=1$ and selected values of temperature T=5, 15, 22 and 100 K. In (b) for $\Delta=0$, T=20 K and selected different values of the exchange coupling (J1: $J_{Rh-Rh}=J_{Rh-Cr}=J_{Cr-Cr}=2$), (J2: $J_{Rh-Rh}=J_{Rh-Cr}=J_{Cr-Cr}=1$), (J3: $J_{Rh-Rh}=J_{Rh-Cr}=-1$ $J_{Cr-Cr}=1$) and (J4: $J_{Rh-Rh}=-1$, $J_{Rh-Cr}=1$ $J_{Cr-Cr}=-1$). In (c) for T=25 K, $J_{Rh-Rh}=J_{Cr-Cr}=J_{Rh-Cr}=1$ and selected values of crystal field $\Delta=-2$, $\Delta=0$, $\Delta=2$ and $\Delta=4$.*

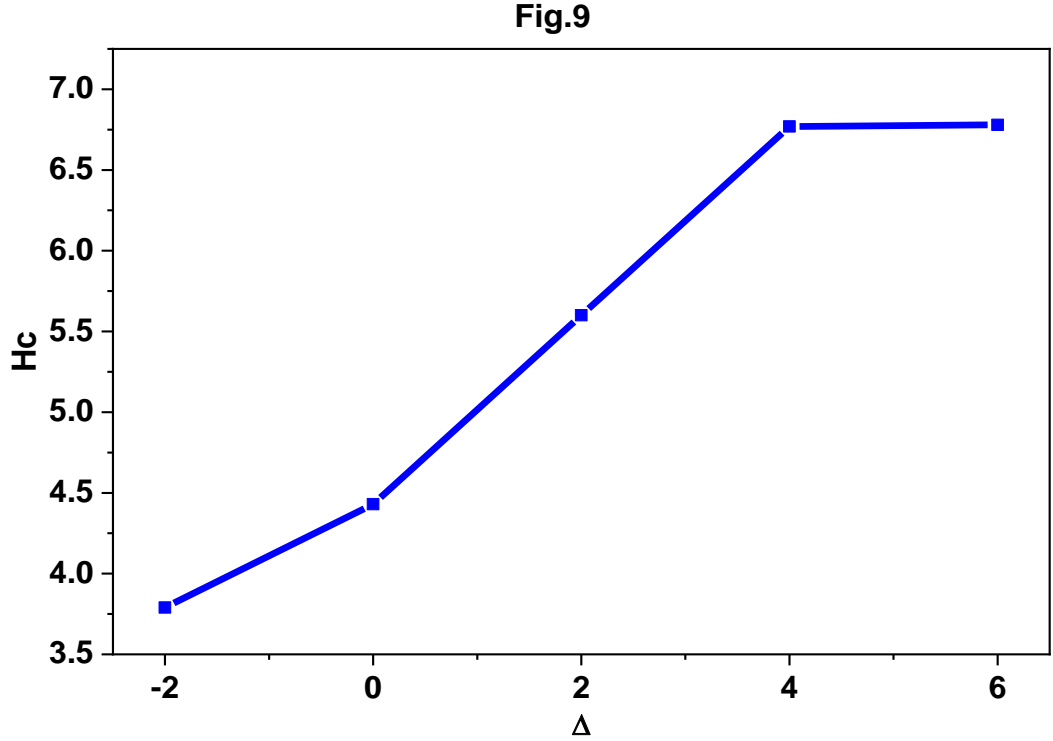

*Fig.9: Coercive field as a function of the crystal field for T=25 K and $J_{Rh-Rh}=J_{Cr-Cr}=J_{Rh-Cr}=1$.*

## IV. Conclusion

In this work, we studied the critical behaviors and magnetic properties of the Half Heusler RhCrSi alloy, using Monte Carlo simulations (MCS) based on the Metropolis algorithm. The studied system contains only two magnetic elements: Rhodium (Rh) and Chromium (Cr). For this design, we have established a model describing the Hamiltonian of the RhCrSi alloy. The Rh and Cr atoms are modeled by the spin moments S = 5/2 and σ=2, respectively. To a primary extent, we compared and searched the ground state phase diagrams in the different physical parameter plans. To illustrate the transition behavior of this arrangement for non-zero temperatures, we examined the total and partial magnetization behaviors using Monte Carlo simulations (MCS). The result of the temperature, the crystal field and the exchange coupling interactions on the total magnetizations have been established and discussed. A detailed example of the behavior of the hysteresis loops of this system was started and discussed for fixed physical parameters.